\documentstyle[psfig,preprint,aps]{revtex}

\begin{document}
\draft
\title{Peripheral heavy ion collisions as a probe of the
nuclear gluon distribution}
 \author{ V.P.
Gon\c{c}alves $^{1,\dag}$\footnotetext{$^{\dag}$
E-mail: barros@ufpel.tche.br} and  C. A. Bertulani
$^{2,\star}$\footnotetext{$^{\star}$
E-mail: bertulani@nscl.msu.edu} }
\address{$^1$ Instituto de F\'{\i}sica e Matem\'atica,  Universidade
Federal de Pelotas\\
Caixa Postal 354, CEP 96010-090, Pelotas, RS, BRAZIL}
\address{$^2$
Department of Physics and Astronomy and National Superconducting Cyclotron
Laboratory, Michigan State University, East Lansing, MI 48824-1321 - USA}
 \maketitle

\begin{abstract}
At  high energies a  quark-gluon plasma is expected to be formed
in heavy ion collisions at RHIC and LHC. The theoretical
description of these processes is directly associated to a
complete knowledge of the details of medium effects in the
nuclear gluon distribution. In this paper we analyze the
possibility to constraint the behavior of this distribution
considering peripheral heavy ion collisions. We reanalyze the
photoproduction of heavy quarks for the deduction of the in-medium
gluon distribution using three current parameterizations for this
parton distribution. Moreover, we show that the elastic
photoproduction of vector mesons is a potential process to  probe
the nuclear gluon distribution.

\end{abstract}


\section{INTRODUCTION}

One of main predictions of QCD is the transition from the
confined/chirally broken phase to the deconfined/chirally
symmetric state of quasi-free quarks and gluons, the so-called
quark-gluon plasma (QGP). Recently the heavy ion collisions have
provided strong evidence for the formation of a QGP \cite{sps},
with the first results of  RHIC marking the beginning of a
collider era in the experiments with relativistic heavy ions, as
well as the  era of detailed studies of  the characteristics of
the QGP. Currently, distinct models associated to different
assumptions describe reasonably the experimental data
\cite{eskqm}, with the main uncertainty present in these analysis
directly connected with the poor knowledge of the initial
conditions of the heavy ion collisions. Theoretically, the early
evolution of these nuclear collisions is governed by the dominant
role of gluons \cite{vni}, due to their large interaction
probability and the large gluonic component in the initial nuclear
wave functions. This leads to a ``hot gluon scenario'', in which
the large number of initially produced energetic partons create a
high temperature, high density plasma of predominantly hot gluons
and a considerably number of quarks. Such extreme condition is
expected to significantly influence QGP signals and should modify
the hard probes produced at early times of the heavy ion
collision. Consequently, a systematic measurement of the nuclear
gluon distribution is of fundamental interest in understanding of
the parton structure of nuclei and to determine the initial
conditions of the QGP. Other important motivation for the
determination of the nuclear gluon distributions is that the high
density effects expected to occur in the high energy limit of QCD
should be manifest in the modification of the gluon dynamics.

At small $x$ and/or large $A$ we expect the transition of the
regime described by the linear dynamics (DGLAP, BFKL) (For a
review, see e.g. Ref. \cite{caldwell}), where only the parton
emissions are considered, to a new regime where the physical
process of recombination of partons becomes important in the
parton cascade and the evolution is given by a nonlinear evolution
equation. In this  regime    a Color Glass Condensate (CGC) is
expected to be formed \cite{mcl}, being characterized by the
limitation on the maximum phase-space parton density that can be
reached in the hadron/nuclear wavefunction (parton saturation) and
very high values of the QCD field strength $F_{\mu \nu} \approx
1/\sqrt{\alpha_s}$ \cite{mue}. In this case, the number of gluons
per unit phase space volume practically saturates and at large
densities grows only very slowly (logarithmically) as a function
of the energy \cite{vicsat}. This implies a large modification of
the gluon distribution if compared with the predictions of the
linear dynamics, which is amplified in nuclear processes
\cite{agl,mclgos,df2a,vicslope}.

Other medium effects are also expected to be present in the
nuclear gluon distribution at large values of $x$: the
antishadowing ($0.1 < x < 0.3$), the EMC effect ($0.3 < x < 0.7$)
and the Fermi motion ($x > 0.7$) \cite{arneodo,weise}. The
presence of these effects is induced from the experimental data
for the nuclear structure function which determines the behavior
of the nuclear quark distributions and the use of the momentum sum
rule as constraint. Experimentally, the behavior of the  nuclear
gluon distribution is indirectly determined in the lepton-nucleus
processes in a small kinematic range of the fixed target
experiments, with the behavior at small $x$ (high energy)
completely undefined. This situation should be improved in the
future with the electron-nucleus colliders at HERA and RHIC
\cite{heraa,raju}, which probably could determine whether parton
distributions saturate. However, until these colliders become
reality we need to consider alternative searches in the current
accelerators which allow us to constraint the nuclear gluon
distribution. Here, we analyze the possibility of using peripheral
heavy ion collisions as a photonuclear collider  and therefore to
determine the behavior of the gluon distribution.

In ultra-peripheral relativistic heavy-ion collisions the ions do
not interact directly with each other and move essentially
undisturbed along the beam direction. The only possible
interaction is due to the long range electromagnetic interaction
and diffractive processes [For a review see, e. g. Ref.
\cite{bert}]. Due to the coherent action of all the protons in the
nucleus, the electromagnetic field is very strong and the
resulting flux of equivalent photons is large. A photon stemming
from the electromagnetic field of one of the two colliding nuclei
can penetrate into the other nucleus and interact with one or more
of its hadrons,   giving rise to photon-nucleus collisions to an
energy region hitherto unexplored experimentally. For example, the
interaction of quasireal photons with protons has been studied
extensively at the electron-proton collider at HERA, with
$\sqrt{s} = 300$ GeV. The obtained $\gamma p$ center of mass
energies extends up to $W_{\gamma p} \approx 200$ GeV, an order of
magnitude larger than those reached by fixed target experiments.
Due to the larger number of photons coming from one of the
colliding nuclei in heavy ion collisions similar and more detailed
studies will be possible in these collisions, with $W_{\gamma N}$
reaching 950 GeV for the Large Hadron Collider (LHC) operating in
its heavy ion mode.

When a  very hard photon from one equivalent swarm of photons
penetrates the other nucleus it is able to resolve the partonic
structure of the nucleus and to interact with the quarks and
gluons. One of the basic process which can occur in the high
energy limit is the photon-gluon fusion leading  to the production
of a quark pair. The main characteristic of this process is that
the cross section is directly proportional to the nuclear gluon
distribution. The analysis of this process in peripheral heavy ion
collisions has been proposed many years ago \cite{perihe1} and
improved in the Refs. \cite{perihe2,perihe3} [For a review, see
Ref. \cite{perihe4}]. Here we reanalyze the charm photoproduction
as a way to estimate the medium effects in $xG_A$ in the full
kinematic region. We consider  as input three distinct
parameterizations of the nuclear gluon distribution. First,  we
consider the possibility that the nuclear gluon distribution is
not modified by medium effects, i.e., $xG_A (x,Q^2) = A \times
xG_N (x,Q^2)$, with the nucleon gluon distribution ($xG_N$) given
by the GRV parameterization \cite{grv95}. Moreover, we consider
that $xG_A (x,Q^2) = R_G \times xG_N (x,Q^2)$, where $R_G$
parameterizes the medium effects as proposed by Eskola, Kolhinen
and Salgado (EKS) \cite{eks}. The main shortcoming of these
parameterizations is that these are based on the DGLAP evolution
equations which are not valid in the small $x$ regime, where the
parton saturation effects should be considered. In order to
analyze the sensitivity of peripheral heavy ion collisions to
these effects we consider as input the parameterization proposed
by Ayala Filho and Gon\c calves (AG) \cite{ag} which improves the
EKS parameterization to include the high density effects. In this
paper we analyze the mass and rapidity distributions of the cross
sections for peripheral Pb + Pb collisions at LHC energies, where
the experimental analysis of this process should be possible. We
conclude that the distinction between the EKS and AG predictions
for the mass distribution is a factor 1.25 in the small mass
region. For the rapidity distribution the difference between the
predictions is larger, which should allow to discriminate between
the results.

One shortcoming of the analysis of  photoproduction of heavy
quarks in peripheral heavy ion collisions to constraint the
nuclear gluon distribution is the linear dependence of the cross
section with this distribution. This implies that only
experimental data with a large statistics and small error will
allow to discriminate the medium effects in the nuclear gluon
distribution. Consequently, it is very important to analyze other
possible processes which have a stronger dependence in $xG_A$.
Here we propose the study of the elastic photoproduction of vector
mesons in peripheral heavy ion collisions as a probe of the
behavior of the nuclear gluon distribution. This process has been
largely studied in $ep$ collisions at HERA, with the perturbative
QCD predictions describing successfully the experimental data
\cite{caldwell}, considering a quadratic dependence of the cross
section with the nucleon gluon distribution. We extend the
formalism used in $ep$ collisions to peripheral heavy ion
collisions, obtaining that the cross section is proportional to
the nuclear gluon distribution squared,  which amplifies the
distinctions associated to medium effects and implies large
differences in the rapidity distribution. Considering the three
distinct parameterizations for the nuclear gluon distribution
described above, we calculate the total cross section for this
process and rapidity distributions for RHIC and LHC energies. Our
results indicate that this process could be used to probe the
nuclear gluon distribution in all kinematic regions, i.e., this
process will be able to estimate the magnitude of the EMC,
antishadowing and high density effects.

This paper is organized as follows. In next section we present a
brief review of the peripheral heavy ion collisions and the main
formulae to describe the photonuclear process in these collisions.
We analyze the photoproduction of heavy quarks in peripheral heavy
ion collisions and present our predictions for the mass and
rapidity distributions. Moreover, we briefly discuss the
contribution of the resolved component of the photon for the
photoproduction of charm quarks. In Section 3 we consider the
elastic photoproduction of vector mesons and analyze the total
cross sections and rapidity distributions for RHIC and LHC
energies. We also present a comparison between this process and
the photoproduction of heavy quarks, which demonstrates that the
elastic photoproduction of vector mesons in peripheral heavy ion
collisions is a potential process to probe the nuclear gluon
distribution. Finally, in Section 4, we present our main
conclusions.

\section{PHOTOPRODUCTION OF HEAVY QUARKS}

At high energies the dominant process occurring when the photon
probes the structure of the nucleus is the photon-gluon fusion
producing a quark pair. For heavy quarks the photoproduction can
be described using perturbative QCD, with the cross section given
in terms of the convolution between the elementary cross section
for the sub-process $\gamma g \rightarrow Q \overline{Q}$ and the
probability of finding a gluon inside the nucleus, i.e., the
nuclear gluon distribution. Basically, the  cross section for
$c\overline{c}$ photoproduction is given by
\begin{eqnarray}
\sigma_{\gamma g\rightarrow c\overline{c}}\, (s) &=&
\int_{2m_c}^{\sqrt s}dM_{c\overline{c}}
    \frac{d\sigma_{c\overline{c}}}{dM_{c\overline{c}}}\, g_A(x,\mu)\,\,,
    \label{sigpho}
\end{eqnarray}
where $d\sigma_{c\overline{c}}/dM_{c\overline{c}}$ is calculable
perturbatively, $M_{c\overline{c}}$
 is the invariant mass of the
$c\overline{c}$ pair with $M^2_{c\overline{c}}={\hat{s}}=xs$, $s$
is the squared CM energy of the $\gamma A$ system, $g_A(x, \mu)$
is the gluon density inside the nuclear medium, $\mu$ is the
factorization scale ($\mu = \sqrt{M^2_{c\overline{c}}}$), and
$m_c$ is the charm quark mass (In this paper we assume that $m_c =
1.45$ GeV).  Moreover, the differential cross section is
\cite{Gluck78}
\begin{eqnarray}
\frac{d\sigma_{\gamma g\rightarrow c\overline{c}}}
{dM_{c\overline{c}}} = \frac {4\pi\alpha\alpha_se_c^2}
{M^2_{c\overline{c}}} \Big[
(1+\epsilon+\frac{1}{2}\epsilon^2)\ln(\frac{1+\sqrt{1-\epsilon}}
{1-\sqrt{1-\epsilon}})  -(1+\epsilon) \sqrt{1-\epsilon}\Big] \,\,,
\label{integ}
\end{eqnarray}
where $e_c$ is the charm charge and
$\epsilon=4m_c^2/M_{c\overline{c}}^2$. From the above expression,
we verify that the cross section is directly proportional to the
nuclear gluon distribution, which implies the possibility  to
constraint its behavior from experimental results for
photoproduction of heavy quarks.

Throughout this publication we use the Born expression for the
elementary photon-gluon cross section [Eq. (\ref{integ})]. QCD
corrections to the Born cross section will not be considered here,
although these corrections modify the normalization of the cross
section by a factor of two. This is justified by the fact that we
are interested in the relative difference between the predictions
of the distinct nuclear gluon distributions, which should be not
modified by the next-to-leading-order corrections.

\begin{figure}[t]
\centerline{\psfig{file=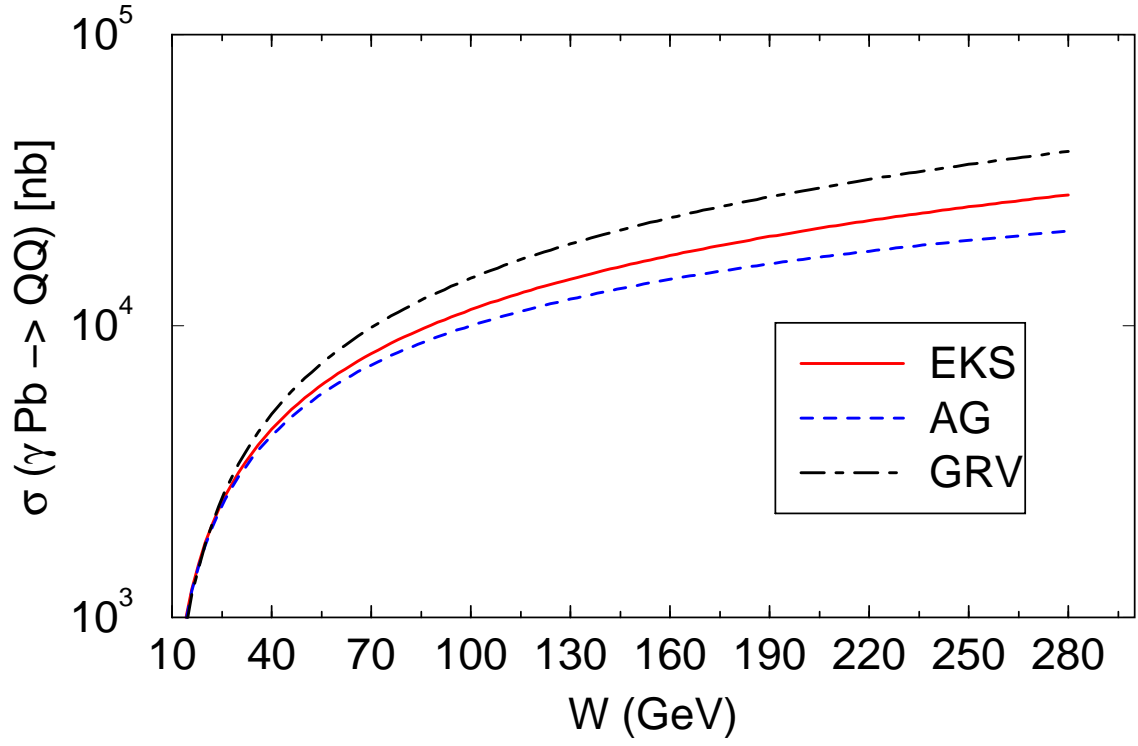,width=150mm}} \caption{Energy
dependence of the photoproduction of heavy quarks for distinct
nuclear gluon distributions ($A=208$).} \label{fig1}
\end{figure}

In Fig. \ref{fig1} we present the energy dependence of the
photoproduction cross section. We focus  our analysis on charm
photoproduction instead of bottom photoproduction, since in this
process smaller values of $x$ are probed.  We verify that
different nuclear gluon distributions imply distinct behaviors for
the cross section, with the difference between the predictions
increasing with the energy. This is associated to the fact that at
high energies we are probing the small $x$ behavior of $xG_A$,
since $x \propto M_{c\overline{c}}/s$, where $M_{c\overline{c}}$
is the invariant mass of the photon-gluon system. Currently, only
the region of small center of mass energy has been analyzed by the
fixed target electron-nucleus experiments, not allowing a good
constraint on medium effects present in the nuclear gluon
distribution. Such situation should be improved in the future with
electron-nucleus colliders at HERA and RHIC \cite{arneodo,raju}.

Another possibility to study photoproduction of heavy quarks at
large center of mass energies is to consider  peripheral heavy ion
collisions \cite{perihe1,perihe2,perihe3}. In this process the
large number of photons coming from one of the colliding nuclei in
heavy ion collisions will allow to study photoproduction, with
$W_{\gamma N}$ reaching  950 GeV for the LHC. To determine the
photoproduction of heavy quarks in peripheral heavy ion collisions
the elementary photon-gluon cross section has to be convoluted
with the photon energy distribution and the gluon distribution
inside the nucleus:
\begin{eqnarray}
\sigma (AA \rightarrow XXQ\overline{Q}) = n(\omega) \otimes
\sigma_{\gamma g \rightarrow Q\overline{Q}} \otimes
xG_A(x,Q^2)\,\,. \label{direct}
\end{eqnarray}
where the photon energy distribution $n(\omega)$ is calculated
within the equivalent photon or Fermi-Weizs\"acker-Williams (FWW)
approximation. In this approximation \cite{bert} the cross section
for a photonuclear reaction in a relativistic heavy ion collision
is given by
\begin{eqnarray}
\sigma = \int \frac{d\omega}{\omega}\,n(\omega)\,\sigma_{\gamma
A}(\omega)\,\,\,,
\end{eqnarray}
where $\sigma_{\gamma A}$ is the on-shell elementary photonuclear
cross section. The appropriate number of equivalent photons,
$n(\omega)$, including nuclear absorption at small impact
parameters, $b$, is given by \cite{BN93}
\begin{eqnarray}
n(\omega)= \frac{2Z^2\alpha\omega^2}{\pi \gamma^2} \int_0^\infty
db \ b \ \left[ K_1^2(\xi) +K_0^2(\xi)/\gamma^2 \right] \, \exp
\left\{ - \sigma_{NN} \int_{-\infty}^\infty dz' \int d^3r
\rho(r)\rho(|{\bf r'-r}|) \right\} \,\,, \label{ene}
\end{eqnarray}
where ${\bf r}' = (b,0,z')$,  $\xi = \omega b/\gamma$,
$\sigma_{NN}$ is the nucleon-nucleon cross section (we use
$\sigma_{NN} = 40$ mb), and $\rho(r)$ is the nuclear density. For
Pb we use a Woods-Saxon density distribution
\begin{eqnarray}
\rho(r) = \frac{\rho_0}{ 1+ \exp \left[ (r-c)/a\right] } \, ,
\label{WS}
\end{eqnarray}
with parameters $c=6.63$ fm, $a=0.549$ fm, and $\rho_0 = 0.16$
fm$^{-3}$. For $b\ll 2c$ the exponential function in Eq.
(\ref{ene}) can be replaced by zero, i.e., for central, or almost
central collisions, the nuclei will surely fragment and the
coherent exchange of virtual photons among them is not part of the
main physics involved.

Before we present our results, it is interesting for our studies
to determine the values of $x$ which will be probe in peripheral
heavy ion collisions. The Bjorken $x$ variable is given by $x =
(M/2p) e^{-y}$, where $M$ is the invariant mass of the
photon-gluon system and $y$ the center of momentum rapidity. For
Pb + Pb collisions at LHC energies the nucleon momentum is equal
to $p=3000$ GeV; hence $x = (M/6000 \, {\rm GeV}) e^{-y}$.
Therefore, the region of small mass and large rapidities probes
directly the small $x$ behavior of the nuclear gluon distribution.
This demonstrates that peripheral heavy ion collisions at LHC
represents a very good tool to determine the behavior of the gluon
distribution in a nuclear medium, and in particular the  low $x$
regime. Conversely, the region of large mass and small rapidities
is directly associated to the region where the EMC and
antishadowing effects are expected to be present. Similarly, for
RHIC energies  ($p = 100$ GeV) the cross section will probe the
region of medium and large values of $x$ ($x > 10^{-2}$). For this
kinematical region, the EKS and AG parameterizations are
identical, which implies that the photoproduction  of heavy quarks
in peripheral heavy ion collisions at RHIC does not allows to
constraint the high density effects. However, the study of this
process at RHIC will be very interesting to determine the presence
or not of the antishadowing and EMC effect in the nuclear gluon
distribution.

Here we restrict our considerations to peripheral Pb + Pb
collisions at LHC energies, with the Lorentz contraction factor
set equal to $\gamma = 3000$, in the lab system. At the reference
system of one of the nuclei, the appropriate Lorentz factor is
$\gamma_c = 2\gamma^2 -1$. In this paper we assume that we know
from which nucleus the gluon originates, although in practice we
cannot distinguish  to which nucleus the gluon belongs. Our
calculations assume that the photon is in the field of the nucleus
coming from negative rapidity. If the photon is emitted by the
target instead of the projectile, the resulting rapidity
distribution will be a mirror image of our distributions around
$y=0$, implying that the total rapidity distribution is the sum of
the curves shown in our figures with its mirror images when both
nuclei emit photons.

\begin{figure}[t]
\centerline{\psfig{file=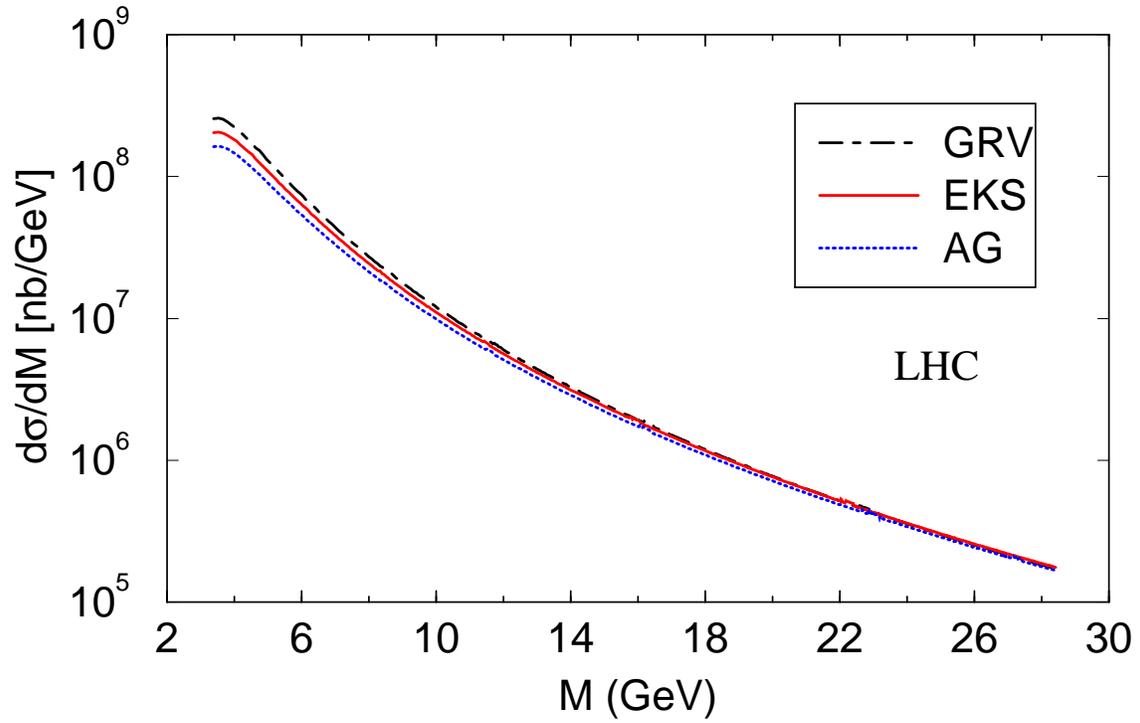,width=150mm}}
\caption{Differential cross section for $c\overline{c}$ production
versus the invariant mass $M=M_{c\overline{c}}$. The photon-gluon
fusion in the heavy-ion collision system $^{208}$Pb + $^{208}$Pb
at LHC energy is considered.} \label{fig2}
\end{figure}

In Fig. \ref{fig2} we present our results for the mass
distribution in photoproduction of charm quarks in peripheral
heavy ion collisions. We can see that the main difference between
the predictions occur at small values of $M$, which is associated
to the small $x$ region. Basically, we have that the predictions
of the EKS parameterization are a factor 1.25 larger than the AG
prediction in this region, while the prediction of Gl\"{u}ck, Reya
and Vogt (GRV) is a factor 2.4 larger. This result is consistent
with the fact that the main differences between the
parameterizations of the nuclear gluon distribution occur at small
$x$ \cite{ag}. The difference between the predictions diminishes
with the growth of the invariant mass, which implies that this
distribution is not a good quantity to estimate the nuclear
effects for medium and large $x$.

\begin{figure}[t]
\centerline{\psfig{file=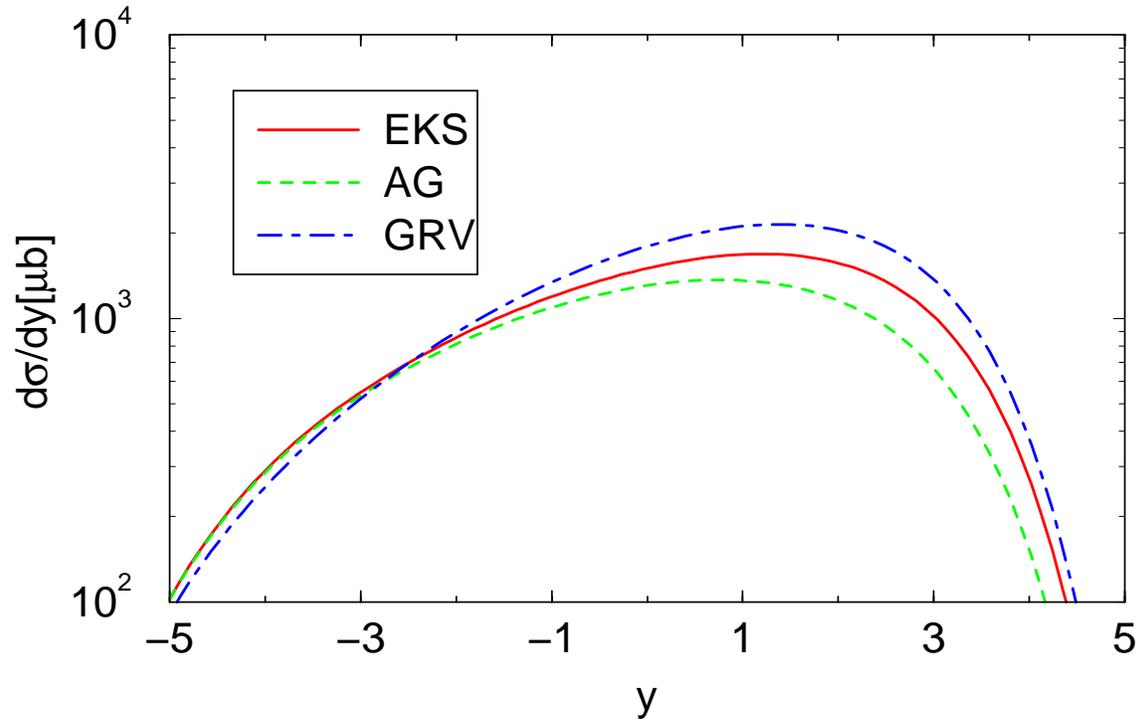,width=150mm}}
\caption{Rapidity distribution for the photoproduction of charm
quarks  in $^{208}$Pb + $^{208}$Pb collisions at  LHC.}
\label{fig3}
\end{figure}

A better distribution to discriminate the behavior of the nuclear
gluon distribution is the rapidity distribution, which is directly
associated to the Bjorken $x$ variable, as discussed above. The
rapidity distribution is calculated considering that $d\sigma/dy =
\omega d\sigma/d\omega$. In Fig. \ref{fig3} we present our results
for the rapidity distribution considering the three
parameterizations of $xG_A$ as input. We have that the region of
small rapidities probes the region of large $x$, while for large
$y$ we directly discriminate the different predictions of $xG_A$
for small $x$. Our results are coherent with this picture: as at
large $x$ the EKS and AG predictions are identical, this region
allows to estimate $R_G = xG_A/(AxG_N)$ in the region of the
antishadowing and EMC effects; at large $y$ the large difference
between the parameterizations implies large modifications in the
rapidity distribution, which should allow a clean  experimental
analysis.

\begin{figure}[t]
\centerline{\psfig{file=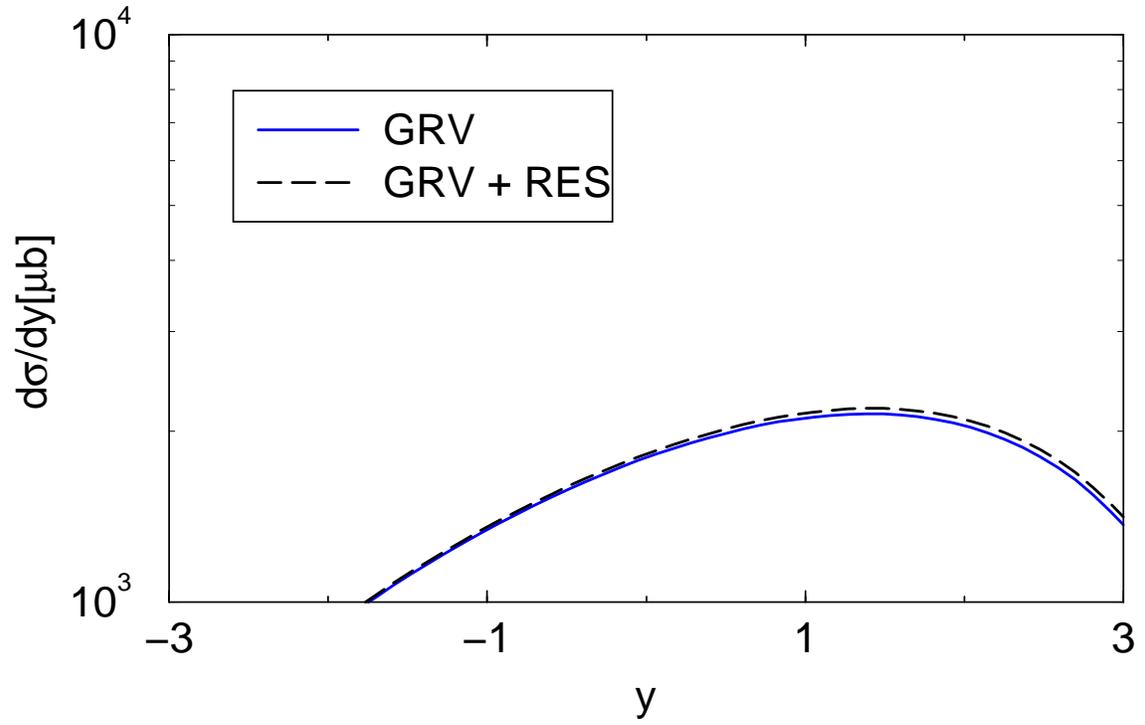,width=150mm}}
\caption{Rapidity distribution for the photoproduction of charm
quarks  in $^{208}$Pb + $^{208}$Pb collisions at  LHC with
(GRV+RES) and without (GRV) the inclusion of the resolved
contribution.} \label{fig3b}
\end{figure}

A comment is in order here. In hard photon-hadron interactions the
photon can behave as a pointlike particle in the so-called direct
photon processes or it can act as a source of partons, which then
scatter against partons in the hadron, in the resolved photon
processes (For a recent review see Ref. \cite{nisius}). Resolved
interactions stem from the photon fluctuation to a quark-antiquark
state or a more complex partonic state, which are embedded in the
definition of the photon structure functions. Recently, the
process of jet production in  photoproduction has been search of
studies of the partonic structure of the photon (See e.g.
\cite{caldwell}), and the contribution of the resolved photon for
the photoproduction of charm has been estimated \cite{frixione}.
Basically, these studies shown that the partonic structure of the
photon is particularly important in some kinematic regions (for
example, the region of large transverse momentum of the charm pair
\cite{frixione}). Consequently, it is important to analyze if this
contribution modifies, for instance, our results for the rapidity
distribution, which is strongly dependent of  nuclear gluon
distribution. In leading order, beyond the process of photon-gluon
fusion considered above, charm production can occur also in
resolved photon interactions, mainly through the process $gg
\rightarrow c\overline{c}$. Therefore, we need to add in Eq.
(\ref{direct}) the resolved contribution  given by
\begin{eqnarray}
\sigma_{res} (AA \rightarrow XXQ\overline{Q}) = n(\omega) \otimes
x_{\gamma}G_{\gamma}(x_{\gamma},Q^2) \otimes \sigma_{g g
\rightarrow Q\overline{Q}} \otimes xG_A(x,Q^2)\,\,,
\label{resolved}
\end{eqnarray}
where $x_{\gamma}$  denote the fraction of the photon momentum
carried by its  gluon component $x_{\gamma} G_{\gamma}$  and
$\sigma_{g g \rightarrow Q\overline{Q}}$ the heavy quark
production cross section first calculated in Ref.
\cite{combridge}. Since the resolved contribution should be the
same for the three nuclear parton distributions,  we only present
in Fig. \ref{fig3b} the results obtained using the  GRV
parameterization for the nucleon. For the photon distribution we
use the GRV photon parameterization \cite{grvphoton}, which
predicts a strong growth of the photon gluon distribution at small
$x_{\gamma}$. We can see that though this contribution be
important in the photoproduction of heavy quarks, as shown in Ref.
\cite{frixione}, it is small in the rapidity distribution of charm
quarks produced in peripheral heavy ion collisions. Therefore, we
believe that the inclusion of the resolved component of the photon
does not   is not relevant for the use of this process as a probe
of the nuclear gluon distribution. Nevertheless, this subject
deserves a further more detailed study.

Before to conclude this section it is important to salient that
the potentiality of the photoproduction of quarks to probe the
high density effects have been recently emphasized in Ref.
\cite{gelis}, where the color glass condensate formalism was used
to estimate the cross section and transverse momentum spectrum.
The authors have verified that the cross section is sensitive to
the saturation scale which characterizes the colored glass. Our
results corroborate the conclusion that this process is sensitive
to the high density effects and the agreement between the
predictions is expected in the kinematic  region in which the
transverse momentum of the pair $k_t$ is larger than the
saturation scale $Q_s$. For $k_t < Q_s$ the collinear
factorization used in this paper to calculate the cross sections
must be generalized, similarly to Ref. \cite{gelis}.

\section{ELASTIC PHOTOPRODUCTION OF VECTOR MESONS}

The production of vector mesons at HERA has become a rich field of
experimental and theoretical research [For a recent review see,
e.g. Ref. \cite{revjpsi}], mainly related with the question of
whether perturbative QCD (pQCD) can provide an accurate
description of  the elastic photoproduction processes. At high
energies the elastic photoproduction of vector mesons is a
two-stage process: at  first the photon fluctuates into the vector
meson which then interacts with the target. For light vector
mesons the latter process occurs similarly to the soft
hadron-hadron interactions and can be interpreted within Regge
phenomenology \cite{dl}. However, at large mass of the vector
meson, for instance the mass of the $J/\Psi$ meson, the process is
hard and pQCD can be applied \cite{brod}. In this case the
lifetime of the quark-antiquark fluctuation is large compared with
the typical interaction time scale and the formation of the vector
meson only occur after the interaction with the target. In pQCD
the interaction of the $q\overline{q}$ pair is described by the
exchange of a color singlet system of gluons (two gluons to
leading order) and, contrary to the Regge approach,  a steep rise
of the vector meson cross section is predicted, driven by the
gluon distribution in the proton. Measurements of the elastic
photoproduction of $J/\Psi$ mesons in $ep$ processes has been
obtained by the H1 and ZEUS collaborations for values of center of
mass energy below 300 GeV, demonstrating the steep rise of  the
cross section predicted by pQCD. This result motivates the
extension of the  pQCD approach used in electron-proton collisions
to photonuclear processes.

The procedure for calculating the forward differential cross
section for photoproduction of a heavy vector meson in the color
dipole approximation is straightforward. The calculation was
performed some years ago to leading logarithmic approximation,
assuming the produced vector-meson quarkonium system to be
nonrelativistic \cite{brod} and improved in distinct aspects
\cite{fran}. To lowest order the $\gamma A \rightarrow J/\Psi A$
amplitude can be factorized into the product of the $\gamma
\rightarrow c \overline{c}$ transition, the scattering of the
$c\overline{c}$ system on the nucleus via (colorless) two-gluon
exchange, and finally the formation of the $J/\Psi$ from the
outgoing $c\overline{c}$ pair.  The heavy meson mass $M_{J/\Psi}$
ensures that pQCD can be applied to photoproduction. The
contribution of pQCD to the imaginary part of the $t=0$
differential cross section of photoproduction of heavy vector
mesons is given by \cite{brod}
\begin{eqnarray}
\frac{d\sigma(\gamma A \rightarrow J/\Psi A)}{dt}|_{t=0} =
\frac{\pi^3 \Gamma_{ee} M_{J/\Psi}^3}{48 \alpha}
\frac{\alpha_s^2(\overline{Q}^2)}{\overline{Q}^8} \times
[xG_A(x,\overline{Q}^2)]^2\,\,, \label{sigela}
\end{eqnarray}
where $xG_A$ is the nuclear gluon distribution, $x =
4\overline{Q}^2/W^2$ with $W$ the center of mass energy and
$\overline{Q}^2 = M_{J/\Psi}^2/4$. Moreover, $\Gamma_{ee}$ is the
leptonic decay width of the vector meson. The total cross section
is obtained by integrating over the momentum transfer $t$,
\begin{eqnarray}
\sigma(\gamma A \rightarrow J/\Psi A) = \frac{d\sigma(\gamma A
\rightarrow J/\Psi A)}{dt}|_{t=0} \, \int_{t_{min}}^{\infty} dt \,
|F(t)|^2 \,\,\,, \label{photonuc}
\end{eqnarray}
where $t_{min} = (M_{J/\Psi}^2/2\omega)^2$ and $F(t)=\int d^3r \
\rho(r) \ \exp (i{\bf q}\cdot {\bf r})$ is the nuclear form factor
for the distribution given by Eq. (\ref{WS}).

A comment is in order here. Although some improvements of the
expression (\ref{sigela})  have been proposed in the literature
\cite{fran}, these modifications do not change the quadratic
dependence on $xG_A$. As our goal in this paper is to analyze the
use of this process to constraint the nuclear gluon distribution,
we will restrict ourselves to the use  of Eq. (\ref{sigela}) in
our studies.

\begin{figure}[t]
\centerline{\psfig{file=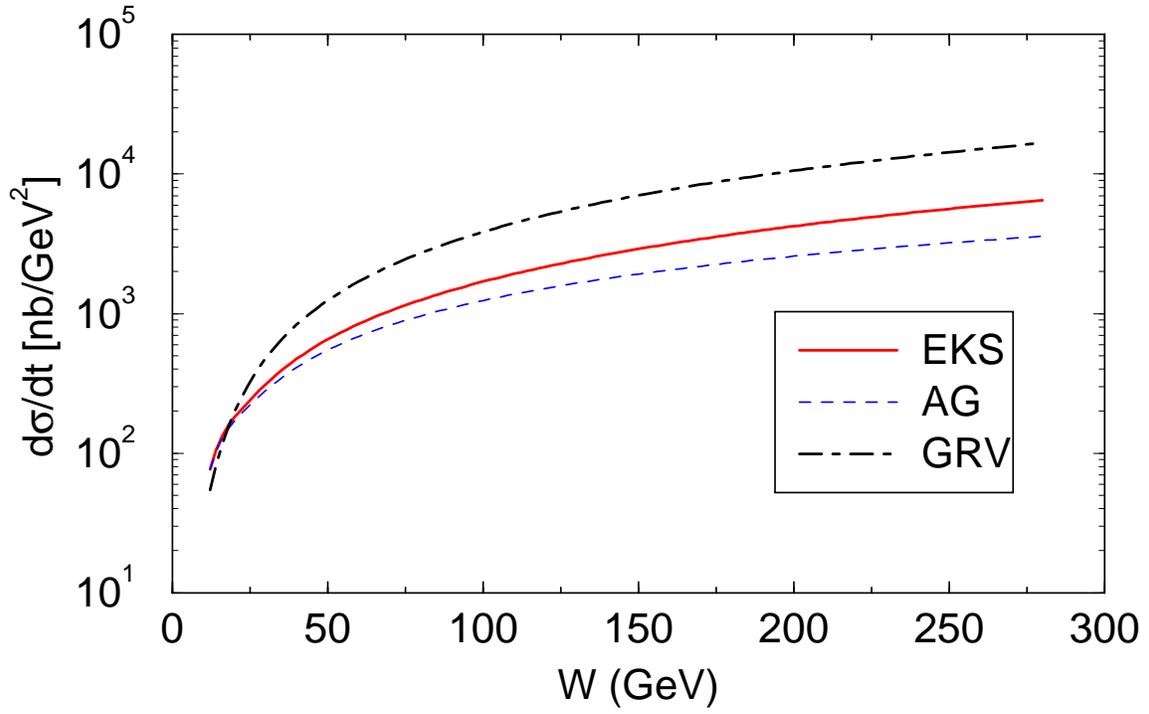,width=150mm}} \caption{Energy
dependence of the elastic photoproduction of $J/\Psi$ for distinct
nuclear gluon distributions ($A=208$).} \label{fig4}
\end{figure}

The main characteristic of the elastic photoproduction of vector
mesons is the quadratic dependence on the gluon distribution,
which makes it an excellent probe of the behavior of this
distribution. In Fig. \ref{fig4} we show the energy dependence of
the differential cross section [Eq. (\ref{sigela})], considering
the three distinct nuclear gluon distributions discussed above. We
have obtained larger differences between the predictions than
obtained in photoproduction of heavy quarks, mainly at large
values of energy. This result motivates experimental analysis of
elastic $J/\Psi$ photoproduction  in photonuclear processes at
high energies. Although the future electron ion colliders (HERA-A
and eRHIC) should probe this kinematic region, here we show that
this can also be done with peripheral heavy ion collisions.

Following similar steps used in photoproduction of heavy quarks,
photons coming from one of the colliding nuclei may interact with
the  other. For elastic photoproduction of $J/\Psi$ we consider
that this photon decays into a $c\overline{c}$ pair which
interacts with the nuclei by the two gluon exchange. After the
interaction, this pair becomes the heavy quarkonium state.
Consequently, the total  cross section for $J/\Psi$ production in
peripheral heavy ion collisions is obtained by integrating the
photonuclear cross section [Eq. (\ref{photonuc})] over the photon
spectrum [Eq. (\ref{ene})], resulting
\begin{eqnarray}
\sigma(AA\rightarrow AAJ/\Psi) = \int \frac{d\omega}{\omega} \
n(\omega) \frac{d\sigma(\gamma A \rightarrow J/\Psi A)}{dt}|_{t=0}
\, \int_{t_{min}}^{\infty} dt \, |F(t)|^2 \,\,.
\end{eqnarray}
In table 1 we present our predictions for the total cross section
considering as input the distinct nuclear gluon distributions and
LHC energies. We salient that although these number will be
modified by the inclusion of higher order corrections for the
cross section \cite{fran}, the difference between the predictions
should not be altered.  We verify that due to the large number of
equivalent photons and the large center of mass energies of the
photon-nucleus system, the cross section for this process is
large, which allows an experimental verification of our
predictions. Also, in peripheral heavy ion collisions the
multiplicity is small what might simplify the experimental
analysis. Moreover, the difference between the predictions is
significant. For RHIC energies, the cross section for this process
is small and probably an experimental determination will be very
hard.

\begin{figure}[t]
\centerline{\psfig{file=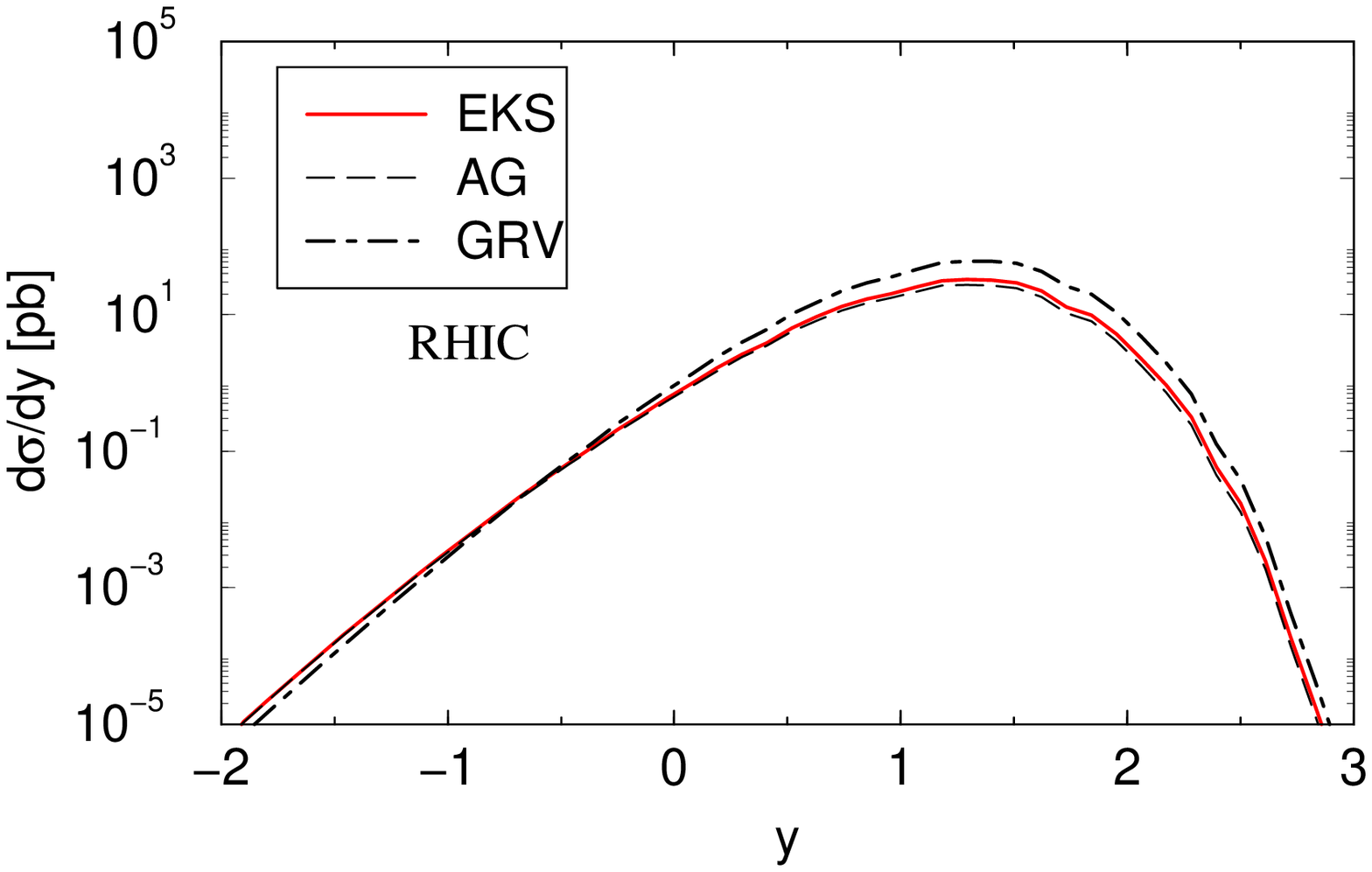,width=150mm}} \caption{The
rapidity distribution for elastic photoproduction of $J/\Psi$ at
RHIC, considering  distinct nuclear gluon distributions
($A=208$).} \label{fig5}
\end{figure}

\begin{figure}[t]
\centerline{\psfig{file=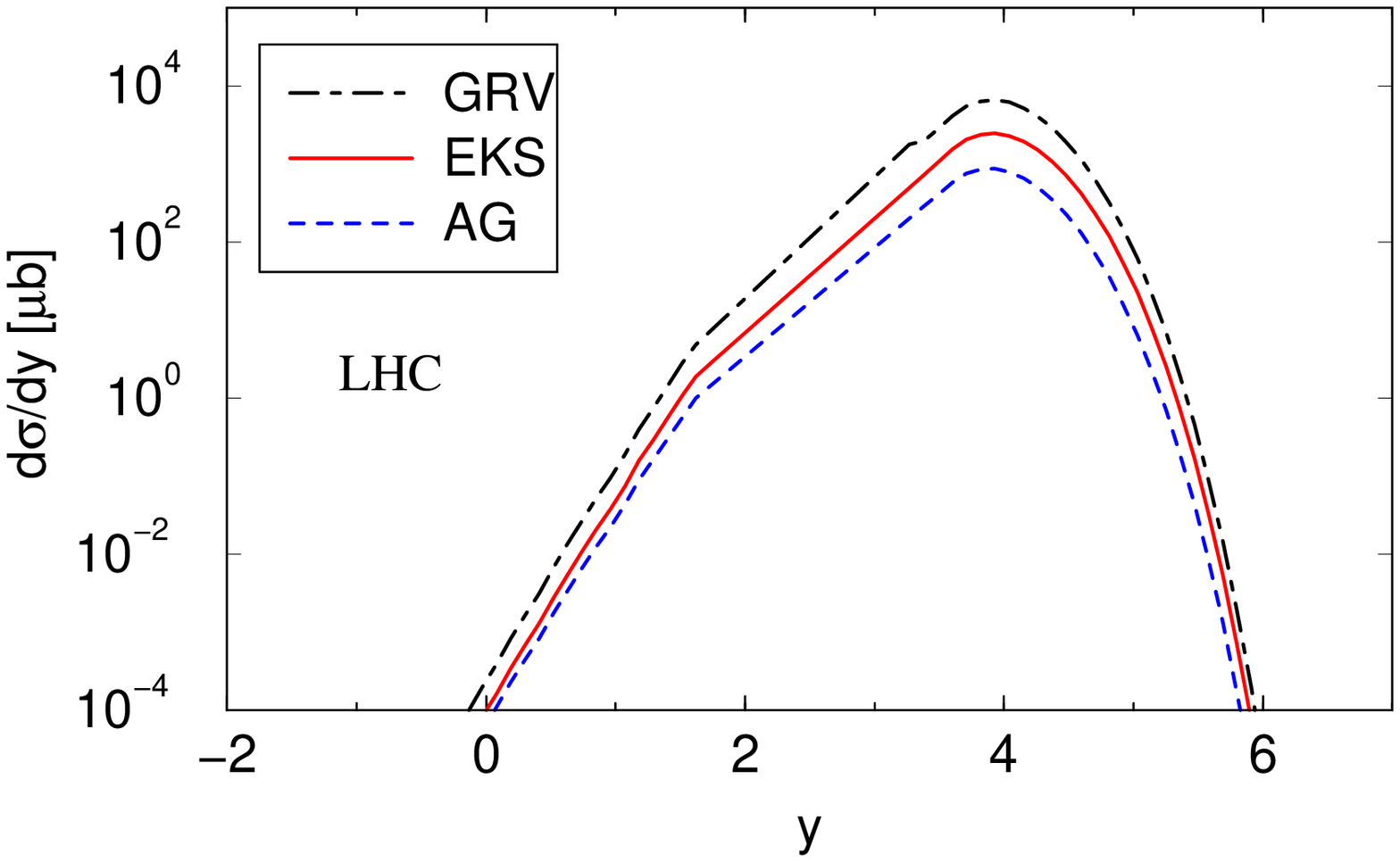,width=150mm}} \caption{The
rapidity distribution for elastic photoproduction of $J/\Psi$ at
LHC considering  distinct nuclear gluon distributions ($A=208$).}
\label{fig6}
\end{figure}

In Figs. \ref{fig5} and \ref{fig6} we present our predictions for
the rapidity distribution for $J/\Psi$ production at RHIC and LHC
energies. In this case, the final state rapidity is determined by
\begin{eqnarray}
y = \frac{1}{2}\ln\frac{\omega}{\sqrt{|t_{min}|}} = \ln \frac{ 2
\omega}{M_{J/\Psi}} \,\,.
\end{eqnarray}
Similarly to the heavy quark photoproduction, the large $y$ region
probes the behavior of $xG_A$ at small $x$, while the region of
small $y$ probes medium values of $x$. We conclude that the
rapidity distribution for  elastic production of $J/\Psi$ at RHIC
allows to discriminate between the GRV and EKS prediction, with
the AG prediction being almost identical to the latter. For LHC we
have a large difference between the distributions, mainly in
magnitude, which  will allow to estimate the magnitude of the EMC,
antishadowing and high density effects.

\begin{figure}[t]
\centerline{\psfig{file=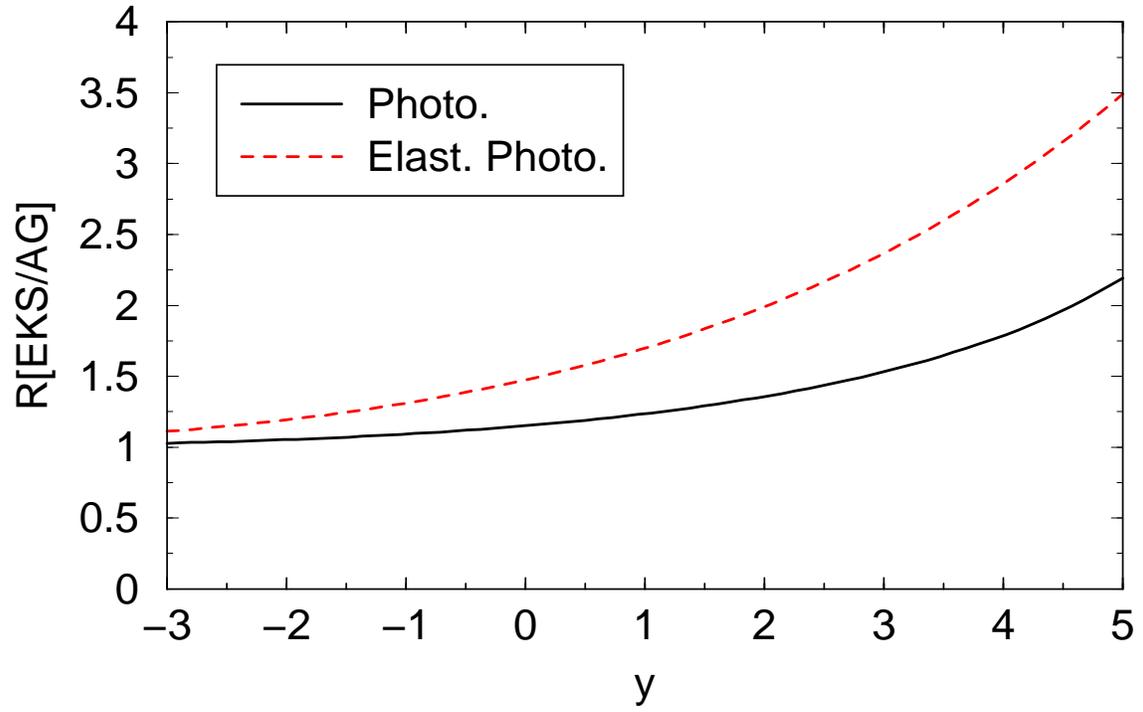,width=150mm}}
\caption{Rapidity behavior of the ratio between the EKS and AG
predictions for photoproduction of charm quarks and elastic
photoproduction of $J/\Psi$.} \label{fig7}
\end{figure}

Finally, in Fig. \ref{fig7} we compare the photoproduction of
heavy quarks and the elastic photoproduction of $J/\Psi$ in
peripheral heavy ion collisions as a possible process to
constraint the behavior of the nuclear gluon distribution. We
present the rapidity distribution of the ratio
\begin{eqnarray}
R[EKS/AG] \equiv \frac{d\sigma}{dy}[EKS]/\frac{d\sigma}{dy}[AG]
\,\,,
\end{eqnarray}
where we consider the EKS and AG parameterizations as inputs of
the rapidity distributions. We confirm that the analysis of the
elastic photoproduction of $J/\Psi$ at medium and large rapidities
is a potential process to determinate the presence and estimate
the magnitude of the high density effects.

\section{CONCLUSIONS}

A systematic determination of gluon distribution is of fundamental
interest in understanding the parton structure of nuclei. The
nuclear gluon distribution is also of central importance in the
field of minijet production that determines the total entropy
produced at RHIC and higher energies. At the moment the behavior
of this distribution is completely undetermined by the fixed
target experiments, with a possible improvement in the future
electron-nucleus colliders. However, as the date of construction
and start of operation of these colliders is still in debate, we
need to obtain  alternative searches to estimate the medium
effects in the nuclear gluon distribution. In this paper  we have
studied the possibility of using  peripheral heavy ion collisions
to constraint the behavior of $xG_A$. In these collisions the high
flux of quasi-real photons from one of the nucleus provides a
copious source of photoproduced reactions and large values for the
cross sections are obtained. Initially, we have reanalyzed the
photoproduction of heavy quarks in peripheral heavy ion collisions
considering three possible parameterizations of $xG_A$ which
estimate the medium effects differently. Our results for the mass
and rapidity distribution shown that at LHC energies this process
is sensitive to the medium effects. However, as this process is
linearly proportional to $xG_A$ the differences are not large,
which implies that only experimental measurements with large
statistics will allow to discriminate between the behaviors.  To
improve this situation, in this paper we propose, for the first
time, the study of the elastic photoproduction of vector mesons in
peripheral heavy ion collisions as a potential process to probe
the medium effects. As the cross section for the elastic vector
meson production depends (quadratically) on the gluon
distribution, it gives a unique opportunity to study the low $x$
behavior of the gluons inside the nucleus. Our results demonstrate
that the study of photoproduction of vector mesons determines the
behavior of the nuclear gluon distribution in the full kinematic
region, with the rapidity distribution allowing for the first time
to estimate the EMC, antishadowing and high density effects.

We expect that our results contribute to motivate the studies of
peripheral heavy ion collisions at RHIC and LHC since, as
discussed in the introduction, the determination of the medium
effects is fundamental for the derivation of reliable predictions
of the initial conditions of the quark gluon plasma and its
signatures.

\section*{ACKNOWLEDGMENTS}
We acknowledge helpful discussions with A. Ayala, F. Gelis, K.
Itakura, J. Jalilian-Marian and R. Venugopalan. VPG thanks the
Brookhaven National Laboratory for its hospitality at a
preliminary stage of this work when this collaboration began. This
work was partially financed by the DOE  under contract No.
DE-AC02-98CH10886, by Brazilian funding agencies CNPq, FAPERGS,
FUJB, and PRONEX, and by the John Simon Guggenheim Foundation
through a fellowship (CAB).

\newpage

\begin{table}[h]
\begin{center}
\begin{tabular} {||l||l||}
\hline \hline Gluon Distribution  & LHC \\
\hline GRV &  $6.584$ mb
\\\hline EKS &  2.452 mb \\
\hline AG &  0.893 mb \\
\hline \hline
\end{tabular}
\end{center}
\label{tab1} \caption{The total cross section $\sigma(AA
\rightarrow AAJ/\Psi)$ for different nuclear gluon distributions.
Results for  LHC.}
\end{table}

\newpage

\newpage

\newpage

\newpage

\newpage

\newpage

\newpage

\newpage

$\,\,\,\,\,\,\,\,\,\,\,\,\,\,\,\,\,\,\,\,\,$

\end{document}